\documentclass{article}
\usepackage{amsfonts}
\usepackage{amsmath}
\usepackage{amssymb}
\usepackage{graphicx}
\usepackage{array}
\usepackage{booktabs}
\usepackage{multirow}
\usepackage{makecell}
\usepackage{float}
\usepackage{booktabs}
\pdfoutput=1 

\usepackage{mymacros}
\usepackage{chngcntr}
\usepackage{verbatim}
\usepackage{amsthm}
\usepackage{graphics}
\usepackage{xcolor}
\usepackage{mathrsfs}
\usepackage{bbold}
\usepackage{cases}
\usepackage{epsfig}
\usepackage{epstopdf}
\usepackage{hyperref}
\usepackage{amsfonts}
\usepackage{hyperref}
\usepackage{jheppub} 

\usepackage[T1]{fontenc} 

\hypersetup{hypertex=true,
	colorlinks=true,
	linkcolor=blue,
	anchorcolor=blue,
	citecolor=blue}
\usepackage[numbers,sort&compress]{natbib}

\title{Topological properties of black holes in five-dimensional gauged supergravity}
\author[a]{Yucheng He \footnote{E-mail: heyucheng365@hotmail.com }}
\author[a]{Changxiang Lei \footnote{E-mail: leichangxiangqqp@outlook.com }}
\author[a]{Deyou Chen \footnote{E-mail: deyouchen@hotmail.com }}

\affiliation{$^{a}$School of Science, Xihua University, Chengdu 610039, China}

\abstract{In this paper, we study the topological properties of five-dimensional rotating charged black holes in different ensembles.The topological numbers for the black holes are gotten in the grand canonical and canonical ensembles, which are 1. When the charge and  cosmological constant disappear, their topological numbers are 0. When the pressure is lower than the critical pressure, the phase transition exists in the canonical ensemble. However, the phase transition also exists in the grand canonical ensemble when the pressure is higher than the critical pressure and two independent rotational parameters are $a=1$ and $b=-1$. This may be due to the fact that the values of the rotational parameters change the supersymmetry of the black hole and lead to the phase transition. }

\begin{document}
	\maketitle

\section{Introduction}

Recently, Wei, Liu and Mann proposed an elegant method to study the black holes. They treated the black hole solutions as defects in the thermodynamic parameter space \cite{WLM}, and used Duan's topological current $\phi$-mapping theory \cite{DL1,DL2} to study the black holes' properties. In this work, the local properties of the black holes are characterized by their winding numbers, and their positive/negative values reflect the local stabilities/unstablities of the black holes. The globe properties are characterized by their topological numbers, and each black hole is endowed with a topological number, therefore, the black holes can be classified by their topological numbers. Inspired by this work, the topological numbers of different black holes were calculated in different ensembles, and the influences of the black holes' parameters such as the angular momentum, charge and cosmological constant on the topological numbers have been extensively discussed \cite{Yerra:2022alz,Yerra:2022coh,Wei:2022mzv,Yerra:2022eov,Bai:2022klw,Fan:2022bsq,Fang:2022rsb,Ye:2023gmk,
Du:2023nkr,Zhang:2023uay,Wu1,Wu2,Wu3,Wu4,Alipour:2023uzo,Sadeghi:2023aii,Ali:2023zww,Wu:2023meo,Zhang:2023tlq,Li:2023ppc,Wei:2023bgp,Li:2023men,Gogoi:2023qku,Hung:2023ggz,Ali:2023jox,Sadeghi:2023dxy,Barzi:2023msl,Chen:2023elp,Du:2023wwg,Rizwan:2023ivp,Liu:2023sbf,Shahzad:2023cis,Bhattacharya:2024bjp,Shahzad:2024ojx,Cunha:2024ajc,Hazarika:2024cpg,Fairoos:2023jvw,Yerra:2023hui}. An interesting phenomenon was found, namely, the topological numbers for some black holes were affected by the ensembles and the values of the black holes' parameters \cite{Liu:2022aqt,Gogoi:2023xzy,Chen:2023ddv,Chen:2024ddv,Hazarika:2023iwp,Gogoi:2023wih}. In view of this, it is natural for people to perform topological classification on black holes in various spacetimes, and to test whether the relevant parameters of black holes affect their classification.

In this paper, we study the topological numbers for the charged rotating black holes in five-dimensional minimal gauged supergravity in the different ensembles, and test whether these numbers are affected by the ensembles and parameters' values for these black holes. These black holes' solution are parameterized by four nontrivial parameters, namely, the mass, charge, and two independent rotation parameters in the two orthogonal $2-$planes, which was obtained in \cite{CCLP}. An interest phenomenon for this gauged supergravity is in the context of the AdS/CFT correspondence, and their bulk properties in anti-de Sitte spacetimes are related to these of strongly coupled conformal field theories on the four-dimensional boundary of AdS$_5$ \cite{Maldacena:1997re,CJY}. Another interest for studying these black holes is that the unfixed cosmological constant can reveal deeper physics and produce many meaningful results. Therefore, it is necessary to study the topological properties of these black holes.

The rest is organized as follows. In the next section, we give a brief review of the solution of the charged rotating black hole in five-dimensional gauged supergravity and discuss its thermodynamic properties. In Sec. \ref{section 3}, we first review the topological approach, and then calculate the topological numbers for the charged rotating black holes and Myers-Perry black holes in canonical and grand canonical ensembles, respectively. Sec. \ref{section 4} is devoted to our conclusion.
	
\section{Five-dimensional black holes in gauged supergravity}

In the five-dimensional gauged supergravity, the general solution of charged rotating black holes in Boyer-Lindquist type coordinates is given by \cite{CCLP}

\begin{eqnarray}
	ds^2=&&-\frac{\Delta_{\theta}[(1+g^2 r^2)\rho^2 dt+2q\nu]dt}{\Xi_{a}\Xi_{b}\rho^{2}}+\frac{2q\nu\omega}{\rho^2}+\frac{f}{\rho^4}\left(\frac{\Delta_{\theta}dt}{\Xi_{a}\Xi_{b}}-\omega\right)^2+\frac{\rho^2dr^2}{\Delta_{r}}\nonumber\\
	&&+\frac{\rho^2d\theta^2}{\Delta_{\theta}}+\frac{r^2+a^2}{\Xi_{a}}\sin^2\theta d\varphi^2++\frac{r^2+b^2}{\Xi_{b}}\cos^2\theta d\psi^2,
	\label{2.1}
\end{eqnarray}

\noindent with an electromagnetic potential

\begin{eqnarray}
A=\frac{\sqrt{3}q}{\rho^2}\left(\frac{\Delta_{\theta}dt}{\Xi_{a}\Xi_{b}}-\omega\right),	
\label{2.2}
\end{eqnarray}

\noindent where

\begin{eqnarray}
	\nu &=& b\sin^2\theta d\varphi+a\cos^2\theta d\psi,\nonumber\\
	\omega&=&a\sin^2\theta \frac{d\varphi}{\Xi_{a}}+b\cos^2\theta \frac{d\psi}{\Xi_{b}},\nonumber\\
	f  &=& 2m\rho^2-q^2+2abqg^2\rho^2, \nonumber\\
	\rho^2  &=& r^2+a^2\cos^2\theta+b^2\sin^2\theta, \nonumber\\
	\Xi_{a}&=&1-a^2g^2,\quad \Xi_{b}=1-b^2g^2,\nonumber\\
	\Delta_{\theta} &=& 1-a^2g^2\cos^2\theta-b^2g^2\sin^2\theta, \nonumber\\
	\Delta_{r} &=&\frac{(r^2+a^2)(r^2+b^2)(1+g^2 r^2)+q^2+abq}{r^2}-2m,
   \label{2.3}
\end{eqnarray}

\noindent $a$ and $b$ are two independent rotation parameters, and their positive/negative values represent the rotational directions. $g$ is a physical quantity expressed as the cosmological constant $g^2=-\frac{\Lambda}{4}$. $m$ and $q$ are parameters related to the ADM mass $M$ and charge $Q$ as follows

\begin{eqnarray}
M &=& \frac{m\pi(2\Xi_{a}+2\Xi_{b}-\Xi_{a}\Xi_{b})+2\pi qabg^2(\Xi_{a}+\Xi_{b})}{4\Xi_{a}^2\Xi_{b}^2},\nonumber\\
Q &=& \frac{\sqrt{3}\pi q}{4\Xi_{a}\Xi_{b}},
\label{2.4}
\end{eqnarray}

\noindent respectively. The metric (\ref{2.1}) describes a Myers-Perry black hole when $g^2=q=0$, and describes a static AdS black hole when $a=b =0$. The electromagnetic potential at the event horizon is

\begin{eqnarray}
\Phi = l^{\mu}A_{\mu}= \frac{\sqrt{3} q r^2_h}{\left(r^2_h+a^2\right) \left(r^2_h+b^2\right)+a b q}.
\label{2.5}
\end{eqnarray}

\noindent where $l^{\mu}= \frac{\partial}{\partial t} + \Omega_a\frac{\partial}{\partial \varphi} +\Omega_b \frac{\partial}{\partial \psi} $ is the Killing vector, and $r_h$ is the event horizon determined by $\Delta_r = 0$. The Hawking temperature and Bekenstein-Hawking entropy are

\begin{eqnarray}
T&=& \frac{r^4_h \left[1+g^2 \left(a^2+b^2+2 r^2_h\right)\right]-(a b+q)^2}{2\pi r_h \left[\left(r^2_h+a^2\right) \left(r^2_h+b^2\right)+a b q\right]},\nonumber\\
S&=& \frac{\pi ^2 \left[\left(r^2_h+a^2\right) \left(r^2_h+b^2\right)+a b q\right]}{2 r_h \Xi_{a}\Xi_{b}}.
\label{2.6}
\end{eqnarray}

\noindent The angular momenta and angular velocities at the horizon are

\begin{eqnarray}
J_a &=& \frac{\pi \left[2ma + qb\left(1+g^2a^2\right)\right]}{4 \Xi_{a}^2\Xi_{b}},\nonumber\\
J_b &=& \frac{\pi \left[2mb + qa\left(1+g^2b^2\right)\right]}{4 \Xi_{a}\Xi_{b}^2},\nonumber\\
\Omega_a &=& \frac{a \left(r^2_h+b^2\right)\left(1+g^2r^2_h\right)+ b q}{\left(r^2_h+a^2\right) \left(r^2_h+b^2\right)+a b q},\nonumber\\
\Omega_b &=& \frac{b \left(r^2_h+a^2\right)\left(1+g^2r^2_h\right)+ a q}{\left(r^2_h+a^2\right) \left(r^2_h+b^2\right)+a b q}.
\label{2.7}
\end{eqnarray}

\noindent The above thermodynamic quantities obey the first law of thermodynamics,

\begin{eqnarray}
d M= TdS + \Omega_a d J_a +\Omega_b d J_b + \Phi dQ,
\label{2.8}
\end{eqnarray}

\noindent where the cosmological constant is fixed. Recently, the black holes' thermodynamics in the extended phase spaces have been researched, and some interesting phenomena, such as Van der Waals-like phase transition, microstructures of black holes, etc,  were found. In these phase spaces, the cosmological constant is regarded as a variable related to pressure $ P= -\frac{\Lambda}{8\pi} = \frac{(n-1)(n-2)}{16\pi l^2}$, where $l$ is the bulk curvature radius and $n$ is a dimension of spacetimes, its conjugate quantity is a thermodynamic volume $V$. Thus the mass $M$ is interpreted as thermodynamic enthalpy instead of internal energy. Considering the varied constant, the first law of this black hole is

\begin{eqnarray}
d M= TdS + \Omega_a d J_a +\Omega_b d J_b + \Phi dQ + VdP.
\label{2.9}
\end{eqnarray}

\noindent The phase transition occurs at $\frac{\partial_{r_h}T}{\partial_{r_h}S}=0$. Using this condition, we get the critical pressure

\begin{eqnarray}
 P_c=-\frac{a b (a b+q)^3 + 3 \left(a^2+b^2\right) (a b+q)^2  r_{h}^2+ (a b+q) (8 a b+5 q)r_{h}^4+  \left(a^2+b^2\right)r_{h}^6-r_{h}^8}{3 a b  \left(a^2+b^2\right) (a b+q)r_{h}^4 +2\pi \left(12 a^2 b^2+a^4+10 a b q+b^4\right)r_{h}^6 + 5 \left(a^2+b^2\right)r_{h}^8+2 r_{h}^{10}}.
\label{2.11}
\end{eqnarray}

\noindent Then the critical horizon radius for the phase transition obeys

\begin{eqnarray}
&&3 a b \left(a^2+b^2\right)^2 (a b+q)^3 +3 \left(a^2+b^2\right) (a b+q)^2 \left(a^4+9 a^2 b^2+b^4+7 a b q\right)r_{h}^2 \nonumber\\
&&+ 2 (a b+q)^2\left(14 a^4+ 43 a^2 b^2+14 b^4+15 a b q\right)r_{h}^4 + \left(a^2+b^2\right) (a b+q) (74 a b+59 q)r_{h}^6\nonumber\\
&&+ 3 \left(a^4+22 a^2 b^2+b^4+30 a b q+10 q^2\right)r_{h}^8 +5 \left(a^2+b^2\right)r_{h}^{10} -2 r_{h}^{12} =0.
\label{2.12}
\end{eqnarray}

\section{Topological numbers for five-dimensional black holes}	\label{section 3}	

\subsection{Review of topological approach}

According to Ref. \cite{WLM}, the generalized free energy is defined by

\begin{eqnarray}
\mathcal{F} = E- \frac{S}{\tau},
\label{3.1.1}
\end{eqnarray}

\noindent where $E$ and $S$ are the energy and entropy of a black hole system, respectively. $\tau$ is a variable and can be seen as the inverse temperature of the cavity enclosing the black hole. This free energy is off-shell except at $\tau = \frac{1}{T}$. To calculate the topological numbers, a vector is constructed as

\begin{eqnarray}
\phi = \left(\frac{\partial \mathcal{F}}{\partial r_h} , -\cot\Theta \csc\Theta\right).
\label{3.1.2}
\end{eqnarray}

\noindent In this vector, $0<r_h<+\infty$ and $0\le\Theta\le\pi$. Zero points of the vector obtained at $\tau = 1/T$ and $\Theta = \pi/2$ correspond to the on-shell black hole solution. Other points are not the solutions of Einstein field equations, thus they are the off-shell states. $\phi^{\Theta}$ diverges at $\Theta =0$ and $\Theta =\pi$, which leads to that the direction of the vector is outward.

Using Duan's $\phi$-mapping topological current theory \cite{DL1,DL2}, we define a topological current

\begin{eqnarray}
j^{\mu} = \frac{1}{2\pi}\varepsilon^{\mu\nu\rho}\varepsilon_{ab}\partial_{\nu}n^a\partial_{\rho}n^b,
\label{3.1.3}
\end{eqnarray}

\noindent where $\mu,\nu,\rho = 0,1,2$, $a,b = 1,2$, $\partial_{\nu} = \frac{\partial}{x^{\nu}}$ and $x^{\nu} = (\tau, r_h, \Theta)$. $\tau$ is seen as a time parameter of the topological defect. $n^a$ is a unit vector defined by $\left( \frac{n^r}{||n||}, \frac{n^{\Theta}}{||n||}\right)$. It is easily to prove the current is conserved. Using the Jacobi tensor $\varepsilon^{ab}J^{\mu}(\frac{\phi}{x}) = \varepsilon^{\mu\nu\rho} \partial_{\nu}{\phi}^a\partial_{\rho}{\phi}^b$ and two-dimensional Laplacian Green function $\Delta_{\phi^a}ln||\phi||=2\pi \delta^2(\phi)$, the current is rewritten as

\begin{eqnarray}
j^{\mu} = \delta^2(\phi)j^{\mu}\left(\frac{\phi}{x}\right),
\label{3.1.4}
\end{eqnarray}

\noindent which is nonzero only when $\phi^a(x^i) = 0$. Then a topological number in a parameter region $\sum$ is calculated as

\begin{eqnarray}
W= \int_{\sum}j^{0}d^2x= \sum_{i=1}^{N}\beta_i\eta_i = \sum_{i=1}^{N}w_i,
\label{3.1.5}
\end{eqnarray}

\noindent where $j^{0} = \sum_{i=1}^{N}\beta_i\eta_i\delta^2(\vec{x}-\vec{z}_i)$ is the density of the current. $\beta_i$ is Hopf index which counts the number of the loops that $\phi^a$ makes in the vector $\phi$ space when $x^{\mu}$ goes around the zero point $z_i$. Clearly, this index is always positive. $\eta_i$=sign$J^0(\phi/x)_{z_i} = \pm 1$ is the Brouwer degree. $w_i$ is the winding number for the $i$-th zero point of the vector in the region and its values is independent on the shape of the region. In the researches, people found the this value is determined by the (un-)stable black holes. In the following, we will study the topological numbers for the five-dimensional black holes.

\subsection{Topological numbers for five-dimensional black holes in canonical ensemble}

Using Eq. (\ref{3.1.1}), we get the generalized free energy

\begin{eqnarray}
\mathcal{F} &=&\frac{\pi  \left(2 \Xi _a+2 \Xi _b-\Xi_{a}\Xi_{b}\right) \left[\left(r^2_h+a^2\right) \left(r^2_h+b^2\right) \left(1+g^2 r^2_h\right)+2 abq+q^2\right]}{8\Xi _a^2 \Xi _b^2 r^2_h} \nonumber\\
&&+ \frac{ \pi qab g^2 \left(\Xi _a+\Xi _b\right)}{2 \Xi _a^2 \Xi _b^2} - \frac{\pi ^2 \left[\left(r^2_h+a^2\right) \left(r^2_h+b^2\right)+a b q\right]}{2 r_h \Xi_{a}\Xi_{b} \tau}.
\label{3.2.1}
\end{eqnarray}

\noindent Thus the components of the vector $\phi$ are gotten as follows

\begin{eqnarray}
	\phi^{r_{h}} &=& - \frac{ \pi\left[\Xi _a \Xi _b -3(\Xi _a +\Xi _b )\right] \left[2 g^2 r^6_h + \left(g^2 \left(a^2+b^2\right)+1\right)r^4_h -(a b+q)^2\right]}{4 r^3_h \Xi _a^2 \Xi _b^2}\\
	&&+\frac{\pi ^2 \left[a^2 b^2+a b q-\left(a^2+b^2\right)r^2_h -3 r^4_h\right]}{2 r^2_h\Xi _a \Xi _b\tau}, \nonumber\\
	\phi ^{\Theta} &=& -\cot\Theta \csc\Theta.
	\label{3.2.2}
\end{eqnarray}

\noindent Zero points are determined by $\phi^{r_h} = 0$ and $\Theta= \pi/2$. We solve it and get

\begin{eqnarray}
	\tau=\frac{2 \pi  r_h \Xi _a \Xi _b\left[a^2 b^2+a b q-\left(a^2+b^2\right)r^2_h -3 r^4_h\right]}{\left[\Xi _a \Xi _b -3(\Xi _a +\Xi _b )\right] \left[2 g^2 r^6_h + \left(g^2 \left(a^2+b^2\right)+1\right)r^4_h -(a b+q)^2\right]}.
	\label{3.2.3}
	\end{eqnarray}

\begin{figure}[h]
	\centering
	\begin{minipage}[t]{0.48\textwidth}
		\centering
		\includegraphics[width=0.6\linewidth]{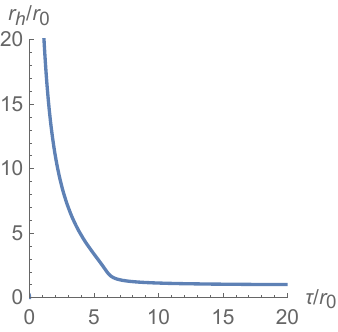}
	\end{minipage}
	\begin{minipage}[t]{0.3\textwidth}
		\centering
		\includegraphics[width=1\linewidth]{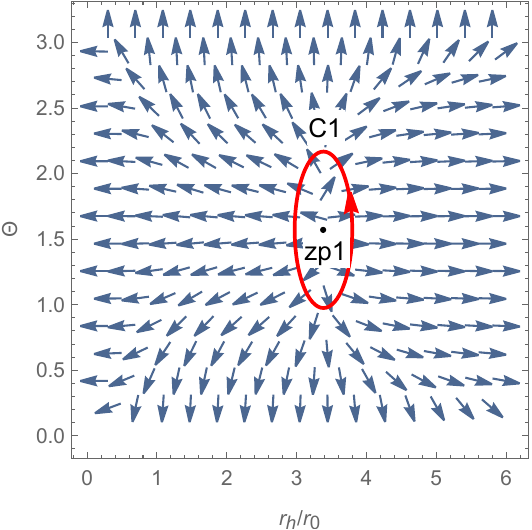}
	\end{minipage}
	\caption{Topological properties of the five-dimensional charged rotating black hole, where $a/r_0 = 0$, $b/r_0 = 1$, $q/r_0^2 = 1$,  $Pr_0^2 = 0.02$ and $P_c r_0^2 = 0.00631$. Zero points of the vector $\phi^{r_h}$ in the plane $r_h - \tau$ are plotted in the left picture. The unit vector field $n$ on a portion of the plane $\Theta - r_h $ at $\tau /r_0=5 $ is plotted in the right picture. The zero point is at $(r_h/r_0 ,\Theta)$= ($3.37, \pi/2$).}		
	\label{fig:3.1.1}
\end{figure}

\begin{figure}[h]
	\centering
	\begin{minipage}[t]{0.48\textwidth}
		\centering
		\includegraphics[width=0.6\linewidth]{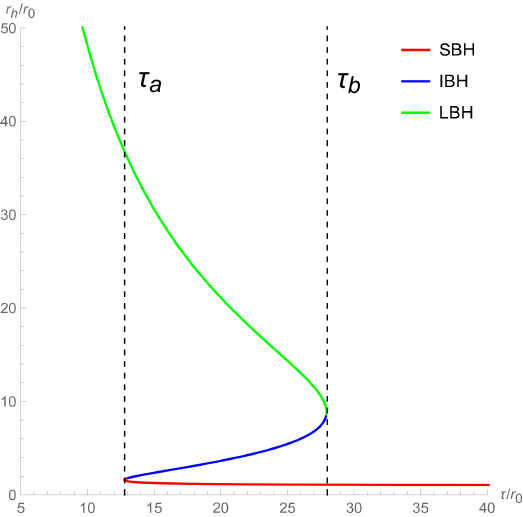}
	\end{minipage}
	\begin{minipage}[t]{0.3\textwidth}
		\centering
		\includegraphics[width=1\linewidth]{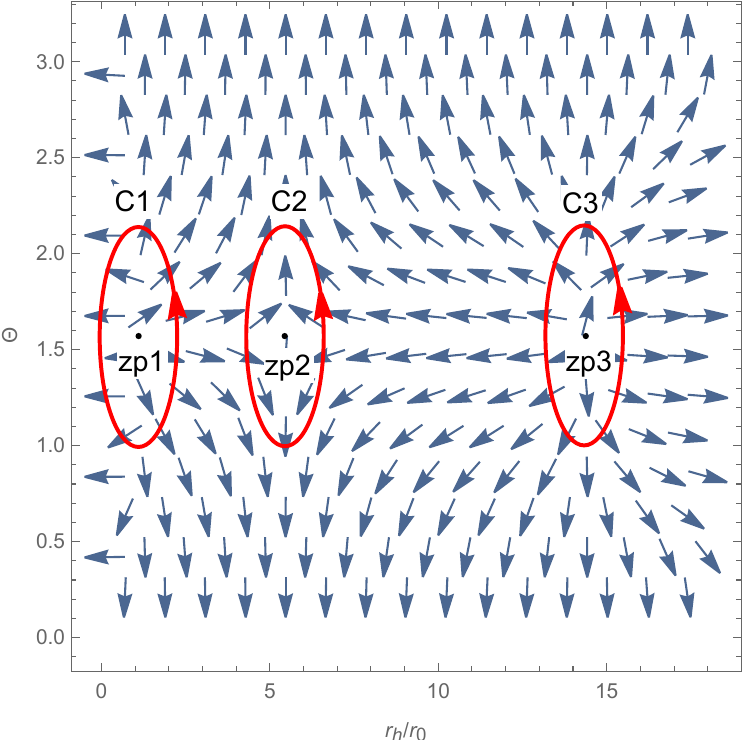}
	\end{minipage}
\caption{Topological properties of the five-dimensional charged rotating black hole, where $a/r_0 = 0$, $b/r_0 = 1$, $q/r_0^2 = 1$,  $Pr_0^2 = 0.001$ and $P_c r_0^2 = 0.00631$. Zero points of the vector $\phi^{r_h}$ in the plane $r_h - \tau$ are plotted in the left picture. The unit vector field $n$ on a portion of the plane $\Theta - r_h $ at $\tau /r_0=25 $ is plotted in the right picture. The zero points are at $(r_h/r_0 ,\Theta)$= ($1.10, \pi/2$), ($5.44, \pi/2$) and ($14.37, \pi/2$), respectively.}			
\label{fig:3.1.2}
\end{figure}

\begin{figure}[H]
	\centering
	\begin{minipage}[t]{0.48\textwidth}
		\centering
		\includegraphics[width=0.6\linewidth]{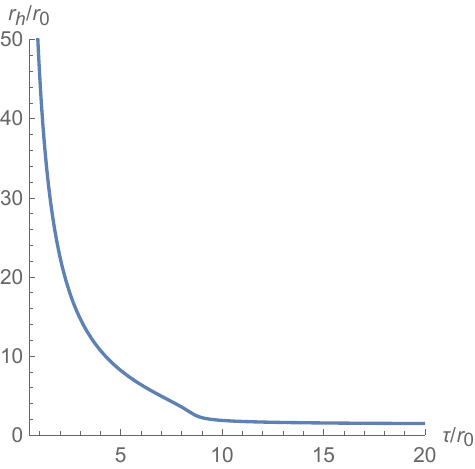}
	\end{minipage}
	\begin{minipage}[t]{0.3\textwidth}
		\centering
		\includegraphics[width=1\linewidth]{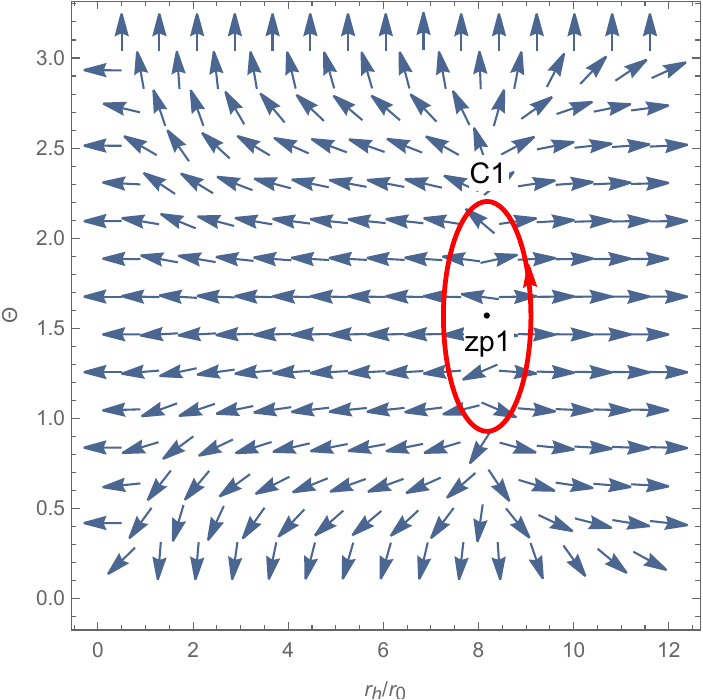}
	\end{minipage}
\caption{Topological properties of the five-dimensional charged rotating black hole, where $a/r_0 = 1$, $ b/r_0 = 1$, $q/r_0^2 = 1$, $Pr_0^2 = 0.01$ and $P_c r_0^2 =0.00286$. Zero points of the vector $\phi^{r_h}$ in the plane $r_h - \tau$ are plotted in the left picture. The unit vector field $n$ on a portion of the plane $\Theta - r_h $ at $\tau /r_0=5 $ is plotted in the right picture. The zero point is at $(r_h/r_0 ,\Theta)$= ($8.17, \pi/2$).}	
\label{fig:3.1.3}
\end{figure}

\begin{figure}[h]
	\centering
	\begin{minipage}[t]{0.48\textwidth}
		\centering
		\includegraphics[width=0.6\linewidth]{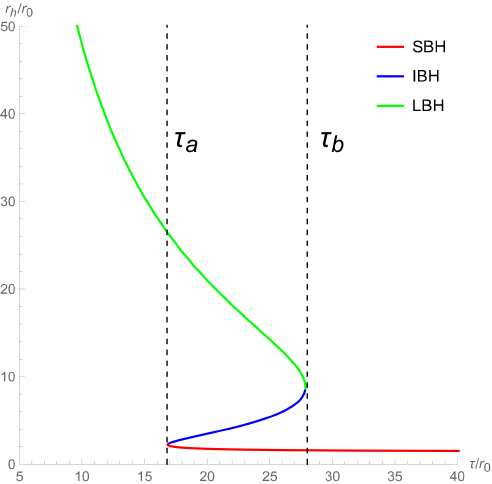}
	\end{minipage}
	\begin{minipage}[t]{0.3\textwidth}
		\centering
		\includegraphics[width=1\linewidth]{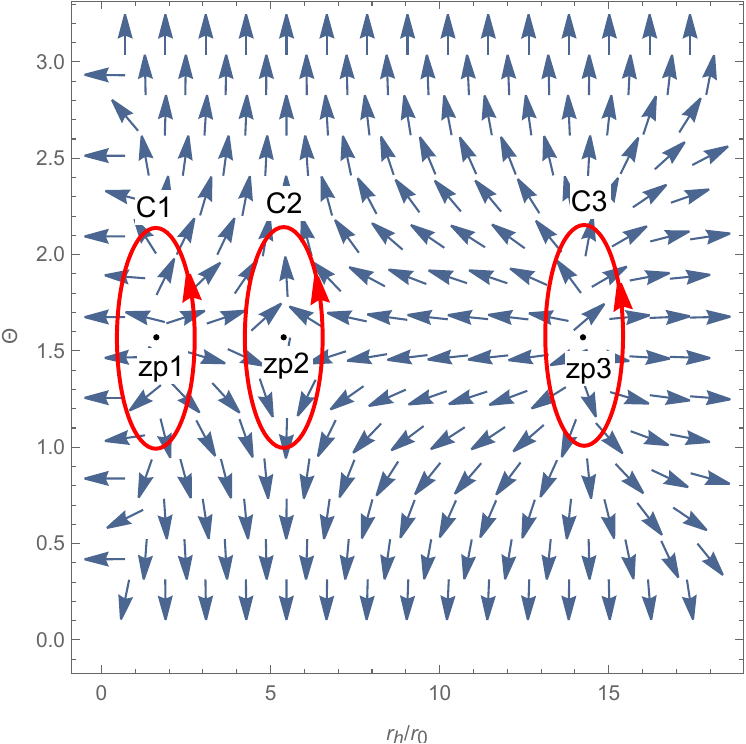}
	\end{minipage}
\caption{Topological properties of the five-dimensional charged rotating black hole, where $a/r_0 = 1$, $ b/r_0 = 1$, $q/r_0^2 = 1$, $Pr_0^2 = 0.001$ and $P_c r_0^2 =0.00286$. Zero points of the vector $\phi^{r_h}$ in the plane $r_h - \tau$ are plotted in the left picture. The unit vector field $n$ on a portion of the plane $\Theta - r_h $ at $\tau /r_0=25 $ is plotted in the right picture. The zero points are at $(r_h/r_0 ,\Theta)$= ($1.62, \pi/2$), ($5.38, \pi/2$) and ($14.23, \pi/2$), respectively.}
\label{fig:3.1.4}
\end{figure}

\begin{figure}[h]
	\centering
	\begin{minipage}[t]{0.48\textwidth}
		\centering
		\includegraphics[width=0.6\linewidth]{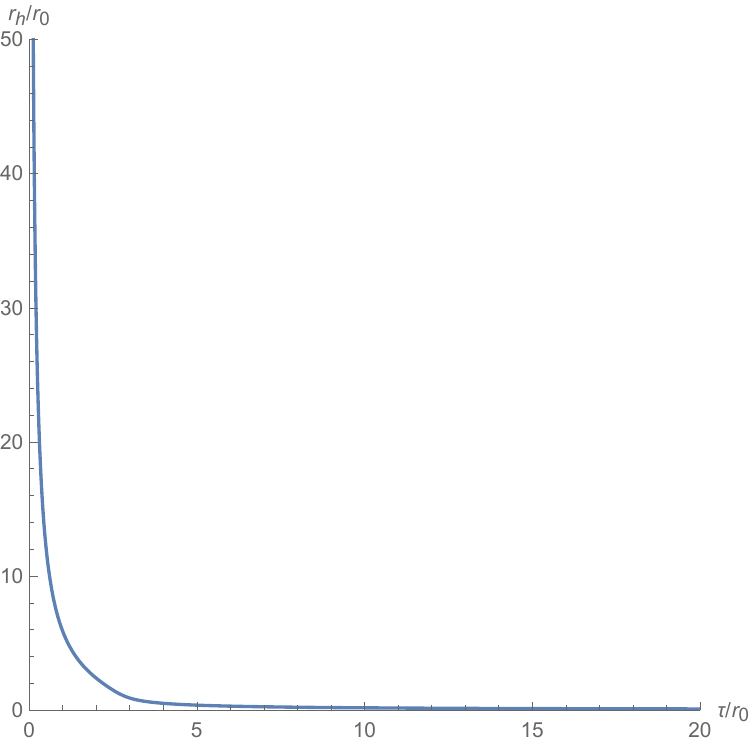}
	\end{minipage}
	\begin{minipage}[t]{0.3\textwidth}
		\centering
		\includegraphics[width=1\linewidth]{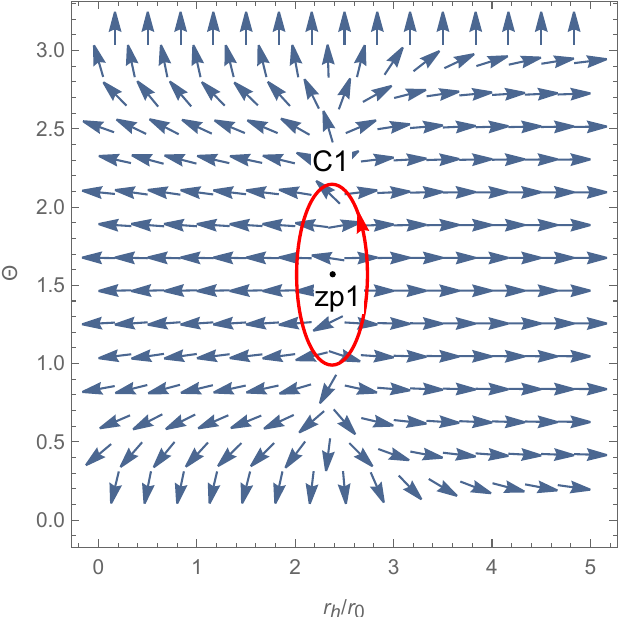}
	\end{minipage}	
\caption{Topological properties of the five-dimensional charged rotating black hole, where $a/r_0 = 1$, $ b/r_0 = -1$, $q/r_0^2 = 1$, $Pr_0^2 = 0.05$ and $P_c r_0^2 =0.00468$. Zero points of the vector $\phi^{r_h}$ in the plane $r_h - \tau$ are plotted in the left picture. The unit vector field $n$ on a portion of the plane $\Theta - r_h $ at $\tau /r_0=2 $ is plotted in the right picture. The zero point is at $(r_h/r_0 ,\Theta)$= ($2.38, \pi/2$).}
\label{fig:3.1.5}
\end{figure}

\begin{figure}[h]
	\centering
	\begin{minipage}[t]{0.48\textwidth}
		\centering
		\includegraphics[width=0.6\linewidth]{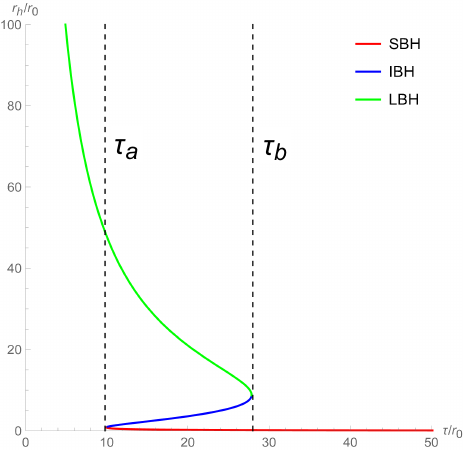}
	\end{minipage}
	\begin{minipage}[t]{0.3\textwidth}
		\centering
		\includegraphics[width=1\linewidth]{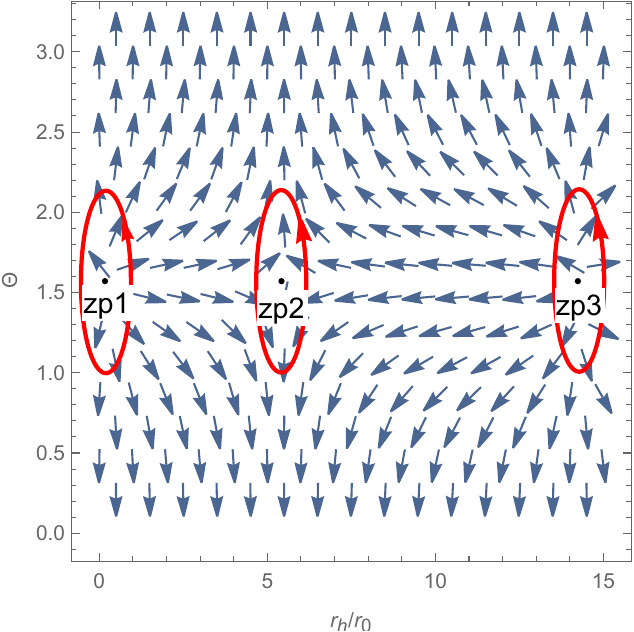}
	\end{minipage}
\caption{Topological properties of the five-dimensional charged rotating black hole, where $a/r_0 = 1$, $ b/r_0 = -1$, $q/r_0^2 = 1$, $Pr_0^2 = 0.001$ and $P_c r_0^2 =0.00468$. Zero points of the vector $\phi^{r_h}$ in the plane $r_h - \tau$ are plotted in the left picture. The unit vector field $n$ on a portion of the plane $\Theta - r_h $ at $\tau /r_0=25$ is plotted in the right picture. The zero points are at $(r_h/r_0 ,\Theta)$= ($0.17, \pi/2$), ($5.42, \pi/2$) and ($14.24, \pi/2$), respectively.}
\label{fig:3.1.6}
\end{figure}

\noindent The above equation shows the relation between the inverse temperature and horizon radius for the zero points. The critical pressure for different values of the parameters is calculated numerically. Topological properties of the black holes are shown in Figure (\ref{fig:3.1.1}) - (\ref{fig:3.1.6}).

In Figure (\ref{fig:3.1.1}) and (\ref{fig:3.1.2}), we first order $a =0$ and study the case where only one rotational parameter exists. When the pressure is larger than the critical pressure we plot in Figure (\ref{fig:3.1.1}). In this figure, the horizon radius decreases monotonically with the increase of the inverse temperature, which shows that the phase transition doesn't exist. Thus the black hole is stable for any temperature. We use the method developed in \cite{WLM} and find that the winding number is $ 1$. There is only one zero point in the right picture of the figure. The winding number can also be calculated by using the method in \cite{Wei1,Wei2}, and it is independent of the loop that surround the zero point. Thus its topological number is $1$. When the pressure is less than the critical pressure, a generation point and an annihilation point exist in Figure (\ref{fig:3.1.2}), and they divide the black hole into three branches: a large black hole (LBH), an intermediate black hole (IBH), and a small black hole (SBH). These three branches coexist in the  range $\tau_a < \tau < \tau_b$, which implies that a LBH/SBH phase transition (PT) exists in this range. For the LBH and SBH branches, they are locally stable, and both have a winding number of $1$. For the IBH branch, it is locally unstable, and its winding number is $-1$. Thus the topological number for this black hole is also $1$.

When two rotational parameters coexist in the black hole, we draw Figure (\ref{fig:3.1.3}) - (\ref{fig:3.1.6}). The positive and negative values of $a$ and $b$ represent the direction of rotation for the black hole on $\varphi$ and $\psi$, respectively. When the pressure is larger than the critical pressure, the radius decreases monotonically with $\tau$ 's value in Figure (\ref{fig:3.1.3}) and (\ref{fig:3.1.5}). When the pressure is less than the critical pressure,   there is also a LBH/SBH phase transition in Figure (\ref{fig:3.1.4}) and (\ref{fig:3.1.6}), which is similar to the case in Figure (\ref{fig:3.1.2}). Comparing to  Figure (\ref{fig:3.1.2}), we find some differences in Figure (\ref{fig:3.1.4}) and (\ref{fig:3.1.6}). Due to the presence of the parameter $a$, $\tau$'s values corresponding to the generation point and annihilation point change, and its range for the PT decrease. All of them show that the topological number for this black hole is $1$, which is fully confirms to the result obtained in \cite{WLM} that the topological number is a universal number independent of the black hole parameters.

\begin{figure}[h]
	\centering
	\begin{minipage}[t]{0.48\textwidth}
		\centering
		\includegraphics[width=0.6\linewidth]{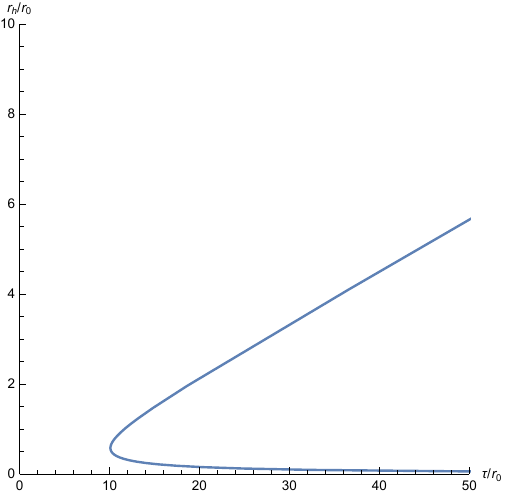}
	\end{minipage}
	\begin{minipage}[t]{0.3\textwidth}
		\centering
		\includegraphics[width=1\linewidth]{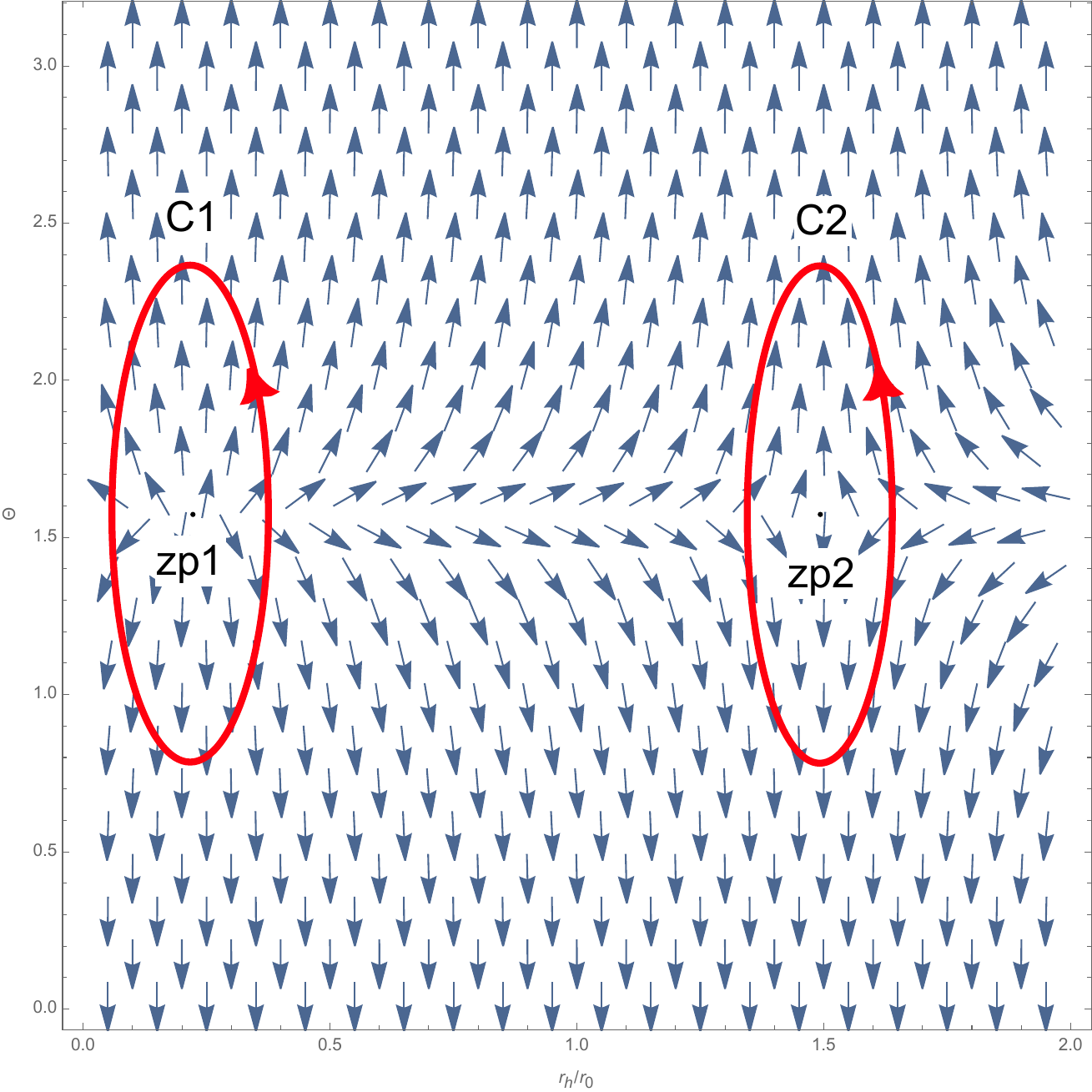}
	\end{minipage}
\caption{Topological properties of the Myers-Perry black hole, where $a/r_0 =1$ and  $b/r_0 =0$. Zero points of the vector $\phi^{r_h}$ in the plane $r_h - \tau$ are plotted in the left picture. The unit vector field $n$ on a portion of the plane $\Theta -r_h $ at $\tau /r_0=25$ is plotted in the right picture. The zero points are at $(r_h/r_0 ,\Theta)$= ($0.21, \pi/2$) and ($1.49, \pi/2$), respectively.}
\label{fig:3.1.7}
\end{figure}

\begin{figure}[h]
	\centering
	\begin{minipage}[t]{0.48\textwidth}
		\centering
		\includegraphics[width=0.6\linewidth]{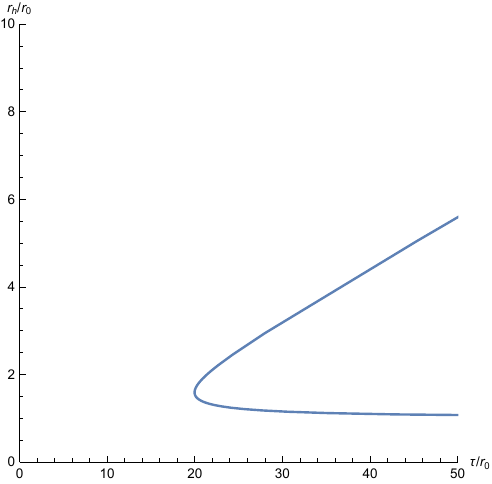}
	\end{minipage}
	\begin{minipage}[t]{0.3\textwidth}
		\centering
		\includegraphics[width=1\linewidth]{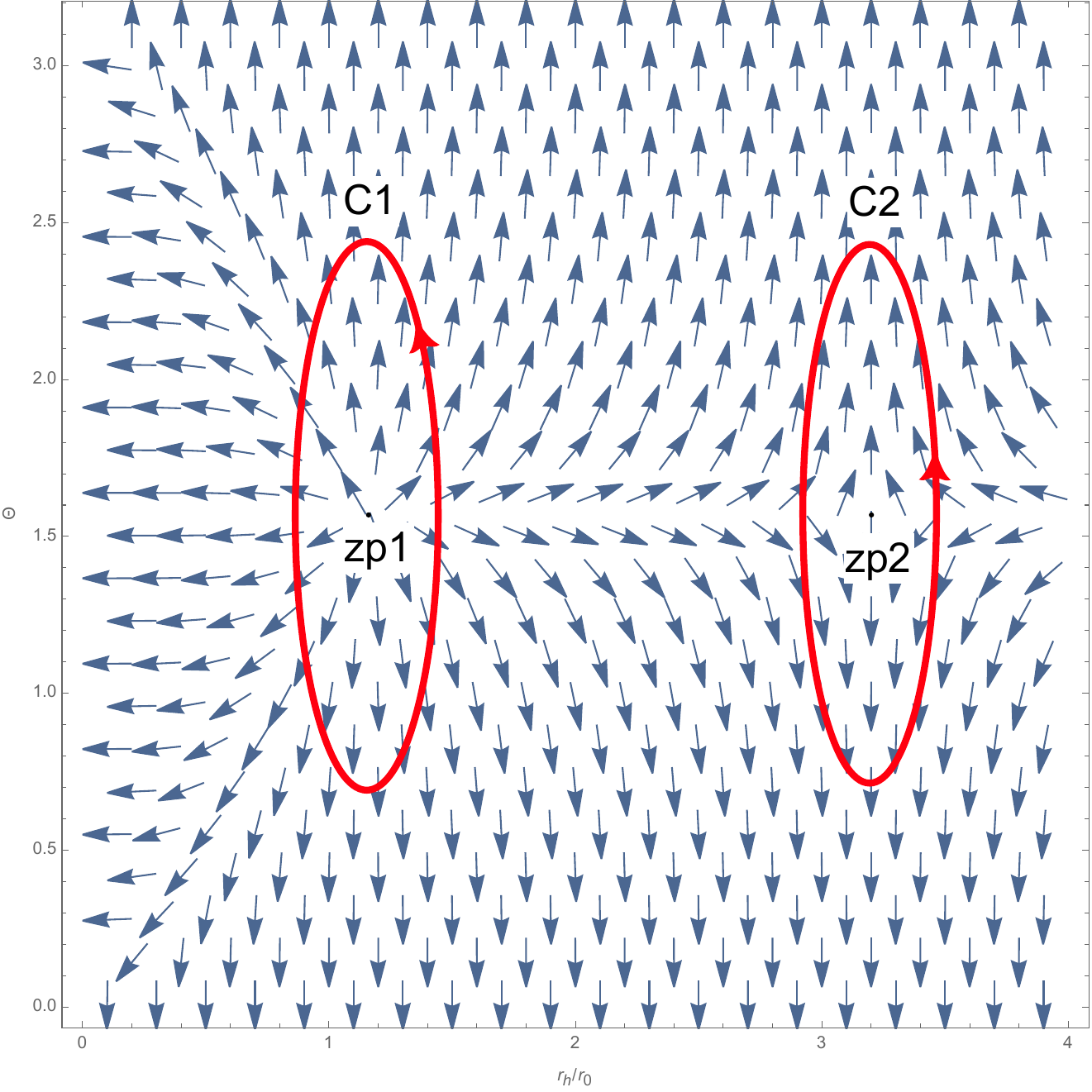}
	\end{minipage}
\caption{Topological properties of the Myers-Perry black hole, where $a/r_0 =1, b/r_0 =1$. Zero points of the vector $\phi^{r_h}$ in the plane $r_h - \tau$ are plotted in the left picture. The unit vector field $n$ on a portion of the plane $\Theta -r_h $  at $\tau /r_0=25$ is plotted in the right picture. The zero points are at $(r_h/r_0 ,\Theta)$= ($1.17, \pi/2$) and ($3.22, \pi/2$), respectively.}
\label{fig:3.1.8}
\end{figure}

The metric (\ref{2.1}) describes a Myers-Perry black hole when $g^2=q=0$. To study the topological number for this black hole, we draw Figure (\ref{fig:3.1.7}) and (\ref{fig:3.1.8}) when $a=1$, $b=0$ and $a=1$, $b=1$, respectively. Due to the vanish of the cosmological constant, we don't need consider the influence of the pressure in the figures. In Figure (\ref{fig:3.1.7}), a generation point divides the black hole into two regions. When $\tau$ 's value exceeds a certain value, there are two different black holes, where the black holes with a small radius being stable and with a large radius being unstable. Their winding numbers are 1 and -1, respectively. Thus the topological number is 0. When $\tau$'s is less than a certain value, there is no black hole existed. Similarly, the topological number in (\ref{fig:3.1.8}) is also 0.

\subsection{Topological numbers for five-dimensional black holes in grand canonical ensemble}

In \cite{Gogoi:2023xzy,Yerra:2023hui,Gogoi:2023wih}, the authors found that the topological class of the four-dimensional dyonic AdS black hole is ensemble dependent. We recalculate the topological numbers in the grand canonical ensemble to verify whether the numbers are consistent with these gotten in the canonical ensemble in this section. In a canonical ensemble, a black hole in a cavity has an energy exchange with the outside, and its temperature, volume and particle number remain be unchanged. In a grand canonical ensemble, the black hole in the cavity exchanges energy and charge with the outside, and its temperature, volume and chemical potential remain be unchanged.

In the grand canonical ensemble, the generalized free energy is

\begin{eqnarray}
\mathcal{F} &=& M-\text{Q$\Phi $}-\frac{S}{\tau } \nonumber\\
&=& \frac{\pi  \left(2 \Xi _a+2 \Xi _b-\Xi_{a}\Xi_{b}\right) \left[\left(r^2_h+a^2\right) \left(r^2_h+b^2\right) \left(1+g^2 r^2_h\right)+2 abq+q^2\right]}{8\Xi _a^2 \Xi _b^2 r^2_h} \nonumber\\
&&+ \frac{ \pi qab g^2 \left(\Xi _a+\Xi _b\right)}{2 \Xi _a^2 \Xi _b^2}-\frac{3 \pi  q^2 r^2_h}{4 \Xi _a \Xi _b \left[\left(r^2_h+a^2\right) \left(r^2_h+b^2\right)+a b q\right]} \nonumber\\
&&- \frac{\pi ^2 \left[\left(r^2_h+a^2\right) \left(r^2_h+b^2\right)+a b q\right]}{2 r_h \Xi_{a}\Xi_{b} \tau}.
\label{3.3.1}
\end{eqnarray}

\noindent Using the definition $\phi^{r_{h}} = \frac{\partial \mathcal{F}}{\partial r_h}$, we get the component of the vector,

\begin{eqnarray}
	\phi^{r_{h}}=&&-\frac{\pi  \left[\Xi _a \Xi _b -3(\Xi _a +\Xi _b )\right] \left[2 g^2 r^6_h+ r^4_h \left(g^2 \left(a^2+b^2\right)+1\right)-(a b+q)^2\right]}{4 r^3_h \Xi _a^2 \Xi _b^2} \nonumber\\
    &&+\frac{3 \pi  q^2 r_h \left[r^4_h-a b (a b+q)\right]}{2 \Xi _a \Xi _b \left[r^4_h + r^2_h \left(a^2+b^2\right)+a b (a b+q)\right]^2}\nonumber\\
    &&+\frac{\pi ^2 \left[a^2 b^2+a b q-r^2_h \left(a^2+b^2\right)-3 r^4_h\right]}{2 r^2_h \Xi _a \Xi _b\tau}.
\label{3.3.2}
\end{eqnarray}

\noindent We solve $\phi^{r_h} = 0$ and obtain the relation

\begin{eqnarray}
	\tau=\frac{2 \pi  r_h \Xi _a \Xi _b \left[a^2 b^2+a b q-r^2_h \left(a^2+b^2\right)-3 r^4_h\right]}{\frac{6 q^2 r^4_h \Xi _a \Xi _b \left[a b (a b+q)-r^4_h\right]}{\left[r^2_h \left(a^2+b^2\right)+a b (a b+q)+r^4_h\right]^2}+\left[\Xi _a \Xi _b -3(\Xi _a +\Xi _b )\right]\left[2 g^2 r^6_h+ r^4_h \left(g^2 \left(a^2+b^2\right)+1\right)-(a b+q)^2\right]}.
\label{3.3.3}
\end{eqnarray}

\begin{figure}[h]
	\centering
	\begin{minipage}[t]{0.48\textwidth}
		\centering
		\includegraphics[width=0.6\linewidth]{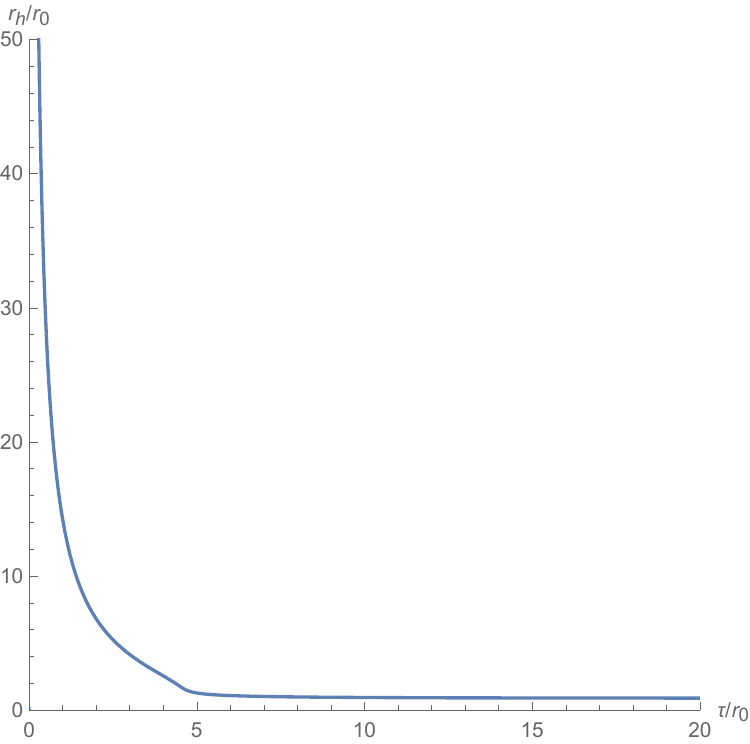}
	\end{minipage}
	\begin{minipage}[t]{0.3\textwidth}
		\centering
		\includegraphics[width=1\linewidth]{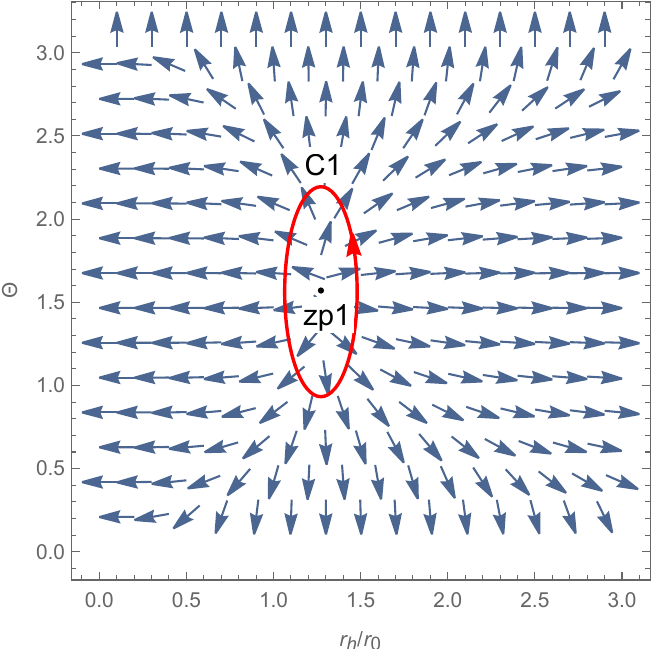}
	\end{minipage}
\caption{Topological properties of the five-dimensional charged rotating black hole in the grand canonical ensemble, where $a/r_0 = 0$, $b/r_0 = 1$, $q/r_0^2 = 1$, $Pr_0^2 = 0.03$ and $P_c r_0^2 =0.00631$. Zero points of the vector $\phi^{r_h}$ in the plane $r_h - \tau$ are plotted in the left picture. The unit vector field $n$ on a portion of the plane $\Theta - r_h$ at $\tau /r_0=5$ is plotted in the right picture. The zero point is at $(r_h/r_0 ,\Theta)$= ($1.27, \pi/2$).}
\label{fig:3.2.1}
\end{figure}

\begin{figure}[h]
	\centering
	\begin{minipage}[t]{0.48\textwidth}
		\centering
		\includegraphics[width=0.6\linewidth]{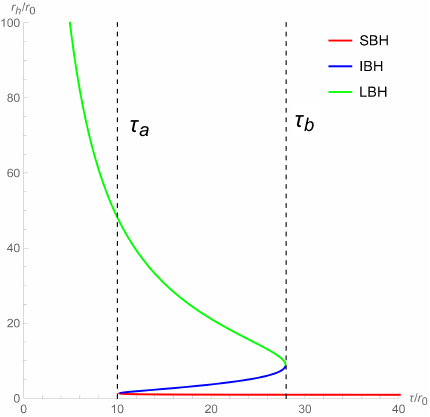}
	\end{minipage}
	\begin{minipage}[t]{0.3\textwidth}
		\centering
		\includegraphics[width=1\linewidth]{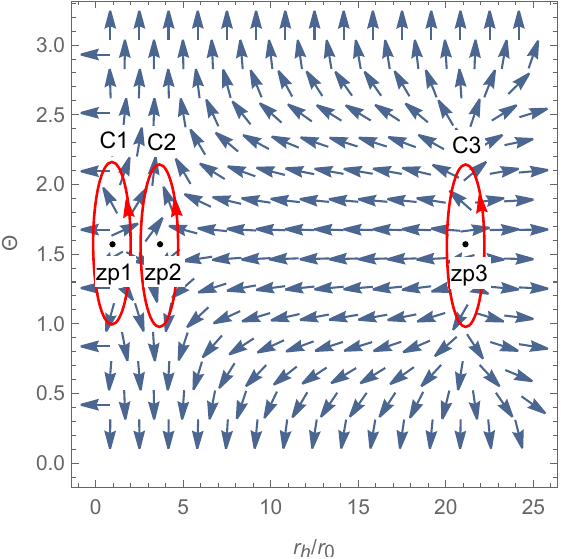}
	\end{minipage}
\caption{Topological properties of the five-dimensional charged rotating black hole in the grand canonical ensemble, where $a/r_0 = 0$, $b/r_0 = 1$, $q/r_0^2 = 1$, $Pr_0^2 = 0.001$ and $P_c r_0^2 =0.00631$. Zero points of the vector $\phi^{r_h}$ in the plane $r_h - \tau$ are plotted in the left picture. The unit vector field $n$ on a portion of the plane $\Theta - r_h$ at $\tau /r_0=20$ is plotted in the right picture. The zero points are at $(r_h/r_0 ,\Theta)$= ($0.96, \pi/2$), ($3.68, \pi/2$) and ($21.12, \pi/2$), respectively.}
\label{fig:3.2.2}
\end{figure}

\begin{figure}[H]
	\centering
	\begin{minipage}[t]{0.48\textwidth}
		\centering
		\includegraphics[width=0.6\linewidth]{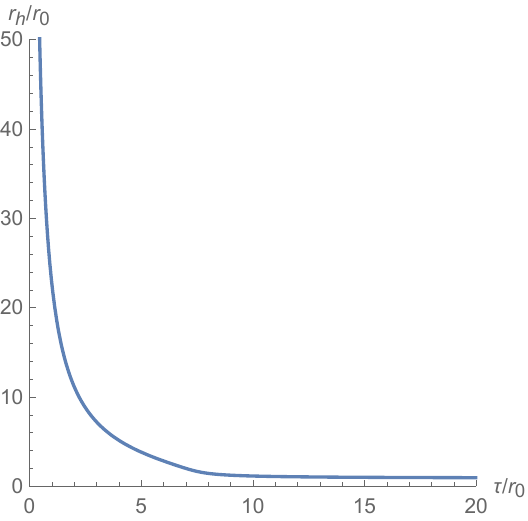}
	\end{minipage}
	\begin{minipage}[t]{0.3\textwidth}
		\centering
		\includegraphics[width=1\linewidth]{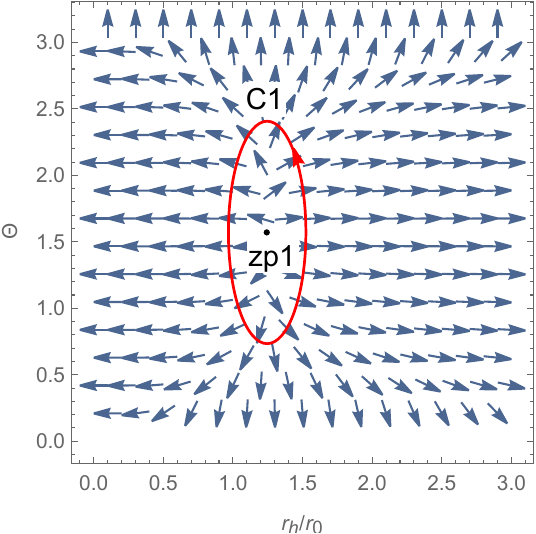}
	\end{minipage}
\caption{Topological properties of the five-dimensional charged rotating black hole in the grand canonical ensemble, where $a/r_0 = 1$, $b/r_0 = 1$, $q/r_0^2 = 1$, $Pr_0^2 = 0.01$ and $P_c r_0^2 =0.00361$. Zero points of the vector $\phi^{r_h}$ in the plane $r_h - \tau$ are plotted in the left picture. The unit vector field $n$ on a portion of the plane $\Theta - r_h$ at $\tau /r_0=5$ is plotted in the right picture. The zero point is at $(r_h/r_0 ,\Theta)$= ($1.26, \pi/2$).}
\label{fig:3.2.3}
\end{figure}

\begin{figure}[H]
	\centering
	\begin{minipage}[t]{0.48\textwidth}
		\centering
		\includegraphics[width=0.6\linewidth]{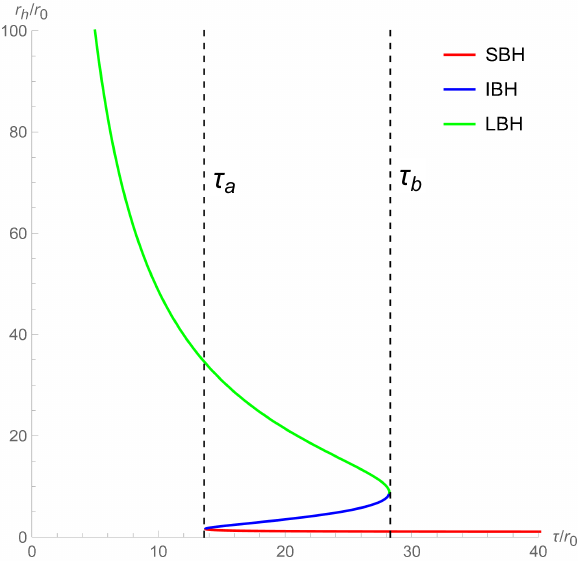}
	\end{minipage}
	\begin{minipage}[t]{0.3\textwidth}
		\centering
		\includegraphics[width=1\linewidth]{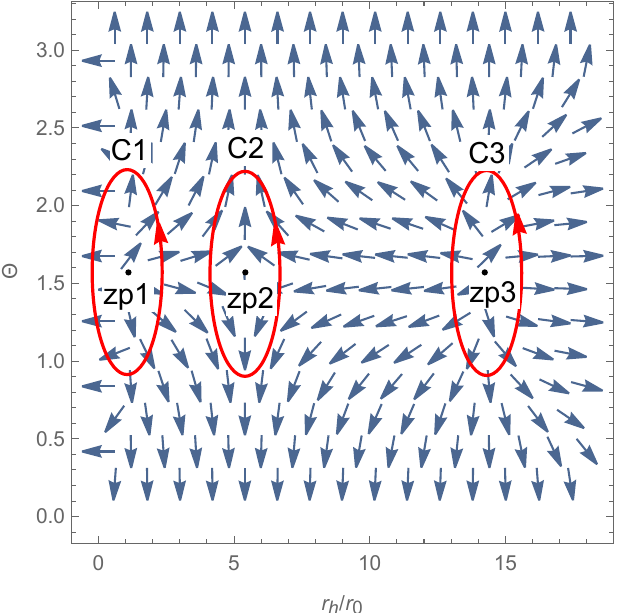}
	\end{minipage}
\caption{Topological properties of the five-dimensional charged rotating black hole in the grand canonical ensemble, where $a/r_0 = 1$, $b/r_0 = 1$, $q/r_0^2 = 1$, $Pr_0^2 = 0.001$ and $P_c r_0^2 =0.00361$. Zero points of the vector $\phi^{r_h}$ in the plane $r_h - \tau$ are plotted in the left picture. The unit vector field $n$ on a portion of the plane $\Theta - r_h$ at $\tau /r_0=25$ is plotted in the right picture. The zero points are at $(r_h/r_0 ,\Theta)$= ($1.12, \pi/2$), ($5.41, \pi/2$) and ($14.22, \pi/2$), respectively.}
\label{fig:3.2.4}
\end{figure}

\begin{figure}[H]
	\centering
	\begin{minipage}[t]{0.48\textwidth}
		\centering
		\includegraphics[width=0.6\linewidth]{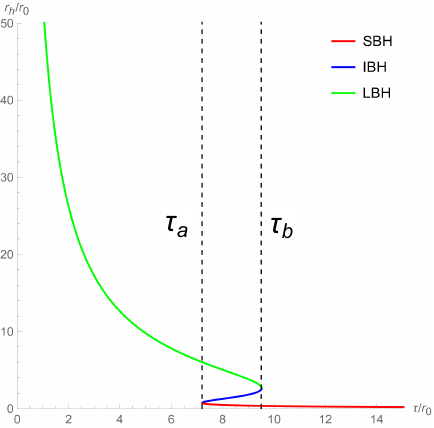}
	\end{minipage}
	\begin{minipage}[t]{0.3\textwidth}
		\centering
		\includegraphics[width=1\linewidth]{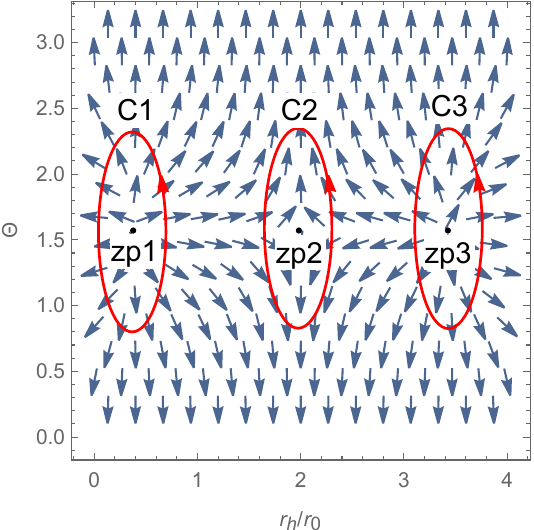}
	\end{minipage}
\caption{Topological properties of the five-dimensional charged rotating black hole in the grand canonical ensemble, where $a/r_0 = 1$, $b/r_0 = -1$, $q/r_0^2 = 1$, $Pr_0^2 = 0.01$ and $P_c r_0^2 =0.00468$. Zero points of the vector $\phi^{r_h}$ in the plane $r_h - \tau$ are plotted in the left picture. The unit vector field $n$ on a portion of the plane $\Theta - r_h$ at $\tau /r_0=8$ is plotted in the right picture. The zero points are at $(r_h/r_0 ,\Theta)$= ($0.38, \pi/2$), ($1.97, \pi/2$) and ($3.43, \pi/2$), respectively.}
\label{fig:3.2.5}
\end{figure}

\begin{figure}[H]
	\centering
	\begin{minipage}[t]{0.48\textwidth}
		\centering
		\includegraphics[width=0.6\linewidth]{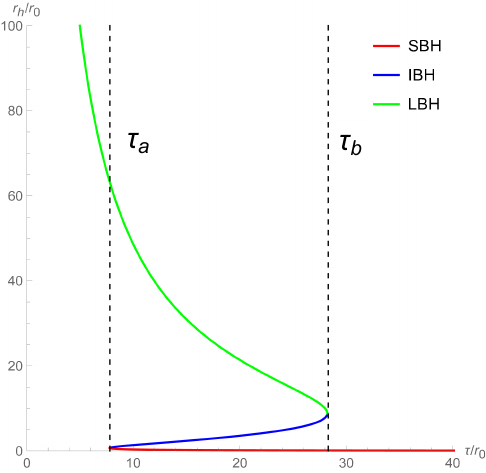}
	\end{minipage}
	\begin{minipage}[t]{0.3\textwidth}
		\centering
		\includegraphics[width=1\linewidth]{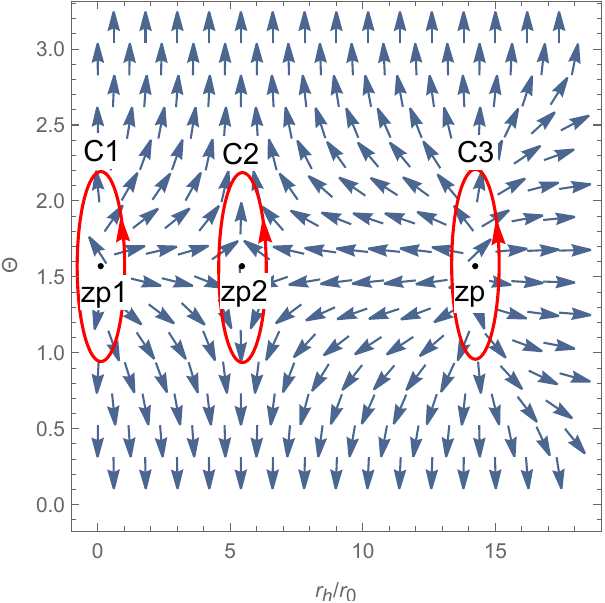}
	\end{minipage}
\caption{Topological properties of the five-dimensional charged rotating black hole in the grand canonical ensemble, where $a/r_0 = 1$, $b/r_0 = -1$, $q/r_0^2 = 1$, $Pr_0^2 = 0.001$ and $P_c r_0^2 =0.00468$. Zero points of the vector $\phi^{r_h}$ in the plane $r_h - \tau$ are plotted in the left picture. The unit vector field $n$ on a portion of the plane $\Theta - r_h$ at $\tau /r_0=25$ is plotted in the right picture. The zero points are at $(r_h/r_0 ,\Theta)$= ($0.10, \pi/2$), ($1.98, \pi/2$) and ($3.42, \pi/2$), respectively.}
\label{fig:3.2.6}
\end{figure}

To study the topological properties in the grand canonical ensemble, we draw Figure (\ref{fig:3.2.1}) - (\ref{fig:3.2.6}). There is only one rotational parameter in Figure (\ref{fig:3.2.1}) and (\ref{fig:3.2.2}). When the pressure is larger than the critical pressure, the horizon radius monotonically decreases with the increase of the temperature, the black hole is local stable and there is no PT. When the pressure is less than the critical pressure, there is a generation point and an annihilation point which divide the black hole into LBH, IBH and SBH branches. The LBH/SBH PT exists in the range $\tau_a < \tau < \tau_b$. These two figures show that the topological number is $1$. When $a= b \ne 0$, the similar result is found in Figure (\ref{fig:3.2.3}) and (\ref{fig:3.2.4}).

We draw Figure (\ref{fig:3.2.5}) and (\ref{fig:3.2.6}) when $b/r_0 = -1$. It is clearly that the topological number is also 1. In the figures, no matter how the pressure is taken, the LBH/SBH PT always exists for this black hole. This strange phenomenon exists in the five-dimensional charged rotating black hole in the grand canonical ensemble. The reason for this phenomenon may be that the values of $a$ and $b$ change the supersymmetry. In \cite{Klemm:2000vn,Gutowski:2004ez}, the authors found that the fraction of supersymmetry is in general $1/4$ for charged rotating black hole solutions in five dimensional gauged supergravity. However, when $a=-b$, the black hole can be viewed as a solution of the pure $N = 2$ gauged supergravity theory. In which case, the preserved supersymmetry becomes $1/2$. When the pressure is higher than the critical pressure, although there is still a LBH/SBH PT, we find that the distance between the generation points and annihilation points in Figure (\ref{fig:3.2.5}) is smaller than that in Figure (\ref{fig:3.2.6}).

\section{Conclusion} \label{section 4}

In this work, we studied the topological properties of the five-dimensional gauged supergravity black holes in the canonical and grand canonical ensembles. The result shows that their topological numbers are 1 in these ensembles. When the charge and  cosmological constant disappear, their topological numbers are 0. This indicates that the cosmological constant and black holes' charges play important roles in the topological classification.

In the work, the cosmological constant was regarded as a variable related to the pressure, and the critical pressure was discussed. We calculated the topological numbers for the black holes when the pressure is greater and less than the critical pressure, respectively. The topological number for the charged rotating black hole is $1$ for any parameters' values. In the canonical ensemble, when the pressure is greater than the critical pressure, there is no generation point and annihilation point. However, when the pressure is less than the critical pressure, there is a generation point and an annihilation point. For the Myers-Perry black hole, since there is no cosmological constant, its topological number is 0, and there is only one generation point. From the above analysis, we conclude that the topological number is always 1 when the black hole is charged and is 0 when the black hole is uncharged. The existence of the PT is related to the critical pressure. However, the LBH/SBH PT exists in the grand canonical ensemble when the pressure is higher than the critical pressure. The reason may be that the values of the rotational parameters change the supersymmetry of the black hole.

\end{document}